\documentclass[preprint]{elsarticle}
\usepackage{graphicx}				
\usepackage{amssymb}
\usepackage{color}

\title{Development of a high brightness ultrafast Transmission Electron Microscope based on a laser-driven cold field emission source}

\author[CEMES]{F. Houdellier\corref{cor}\fnref{http://www.cemes.fr/}}
\ead{florent@cemes.fr}

\author[CEMES]{G.M.~Caruso}

\author[CEMES]{S. Weber}

\author[LPS]{M. Kociak}

\author[CEMES]{A. Arbouet\corref{cor}\fnref{http://www.cemes.fr/}}
\ead{arbouet@cemes..fr}

\address[CEMES]{CEMES-CNRS, Universit\'{e} de Toulouse, Toulouse, France}
\address[LPS]{Laboratoire de Physique des Solides, Bâtiment 510, UMR CNRS 8502, Universit\'{e} Paris Sud, Orsay 91400, France}

\begin{document}

\begin{abstract}
We report on the development of an ultrafast Transmission Electron Microscope based on a cold field emission source which can operate in either DC or ultrafast mode. 
Electron emission from a tungsten nanotip is triggered by femtosecond laser pulses which are tightly focused by optical components integrated inside a cold field emission source close to the cathode.
The properties of the electron probe (brightness, angular current density, stability) are quantitatively determined.
The measured brightness is the largest reported so far for UTEMs.
Examples of imaging, diffraction and spectroscopy using ultrashort electron pulses are given.
Finally, the potential of this instrument is illustrated by performing electron holography in the off-axis configuration using ultrashort electron pulses.
\end{abstract}
\begin{keyword}
Ultrafast Transmission Electron Microscopy, field emission, nanoemitters, electron interferometry,
\end{keyword}
\maketitle

\section{Introduction}
Since its invention in 1931, Transmission Electron Microscopy (TEM) has allowed giant steps in our fundamental understanding of many fields of science including chemistry, biology or physics. Many quantitative techniques have been developed to get structural and chemical information either from imaging (direct space), diffraction (reciprocal space) or spectral analysis (energy space). 
Three types of electron sources can be found in TEMs: thermionic, Schottky and cold-field emission sources. In a thermionic electron source, the electrons are pulled out of a cathode by thermal excitation produced by a strong electric current. In Cold-Field Emission (CFE) electron sources, no heating of the cathode is required. An extraction voltage is instead applied on a sharp conical metallic tip. This voltage, enhanced by the lightning rod effect at the tip apex lowers the potential in vacuum enough to allow for efficient tunneling of electrons out of the tip  from the Fermi level. The very small size of the emitting zone is at the origin of the high brightness of field-emission electron sources. 
The development of field emission guns has allowed TEM to enter in a new era\cite{Crewe1968}. Following the initial suggestion of Gabor in 1948, new techniques such as the off-axis electron holography have started to exploit the high spatial coherence of the electron beam to produce interferograms from which modifications of the phase of the electronic wavepacket can be retrieved\cite{Gabor1948}. These studies have shown that the phase of the electron wave is a very sensitive probe of the electrostatic field, strain field or magnetic field which allows the quantitative mapping of these observables with nanometer resolution \cite{Tonomura1992,Lehmann1999,Dunin2003,Dunin-Borkowski2015}. The brightness of the source, defined as the current per unit area and solid angle, is the figure of merit that must be optimized in order to get the highest spatial coherence or the highest spatial resolution for applications like holography or spatially resolved spectroscopies (in this latter case, the improved energy resolution of the cold FE sources is also very important). 
Schottky guns are halfway and use a field assisted thermionic emission process\cite{swanson1997handbook}. 
They are equipped with a so-called suppressor anode to confine the region from which electrons are emitted to the apex of the tip.
For the most demanding applications, CFEGs are the first choice as they exhibit the highest brightness and the lowest energy spread thanks to their unique combination of small virtual source size and low angular current density.

%
Pioneering work at the Berlin Technical University \cite{Bostanjoglo2000,Domer2003} and Lawrence Livermore National Laboratory \cite{King2007} have clearly established time-resolved Transmission Electron Microscopy as one of the most active line of instrumental developments in TEM.
High speed Transmission Electron Microscopes or Dynamical Transmission Electron Microscopes have provided a unique insight into irreversible processes such as phase transitions, melting or ablation for instance\cite{BOSTANJOGLO1993,BOSTANJOGLO1994}.
Their spatio-temporal resolution was however limited by the large number of electrons in each pulse \cite{Gahlmann2008}.
By using pulses containing only a few electrons, the group of A. Zewail at Caltech has overcome this limitation and performed time-resolved studies with both nanometer spatial resolution and sub-picosecond temporal resolution\cite{Lobastov2005,Zewailbook}. 
Until recently, UTEMs were all based on flat photocathodes implemented in thermionic electron guns.
The large size of the illuminated area on the photocathode was limiting the brightness of these electron sources and prevented their use for the most demanding TEM applications such as electron holography.

A decade ago, it has been shown that laser-driven nanoemitters could provide an exciting alternative to conventional photocathodes\cite{Hommelhoff2006a,Ropers2007}.
Due to the enhancement of the laser electric field at the apex, it is possible to trigger electron emission from a small region and investigate light-matter interaction in strong optical fields\cite{Schenk2010,Bormann2010,Kruger2011,Sivis2013}.
Processes such as electron rescattering which are at the heart of attosecond physics can be investigated with low power high repetition rate laser systems whereas similar studies on dilute systems demand amplified lasers\cite{Krausz2009}.
The confinement of the emission region to strong field regions at the tip apex further yields a brightness in laser-driven mode which is similar to conventional DC modes and makes laser-driven nanoemitters very promising for ultrafast coherent electron microscopies \cite{Ehberger2015}.
The first implementation of a laser-driven field-emission tip in a TEM has been achieved in G\"{o}ttingen on a Schottky-type electron source\cite{Paarmann2012,Bormann2015}.
The spectacular improvement in brightness achieved has allowed to perform unique experiments which would not have been possible on the previous generation of UTEMs\cite{Feist2015,Echternkamp2016,Feist2017}.
The development of an ultrafast electron source based on a cold-field emission gun has been recently demonstrated but
its potential for ultrafast TEM has not been explored yet \cite{Caruso2017}.

It is the purpose of this paper to describe the potential of such an ultrafast Transmission Electron Microscope based on a modified cold field emission source. 
We here follow a different line from the development achieved in G\"{o}ttingen.
The tight arrangement of Schottky sources makes the integration of short focal distance optics inside the electron source difficult.
In this case, the optics used to focus the laser beam has therefore been placed outside the TEM column and the region from which electrons are emitted is restricted by the use of the additional suppressor electrode available on Schottky electron guns and/or by chemical selectivity using a zirconia wetting layer on the [100]  oriented front facet of tungsten tips \cite{swanson1997handbook,Feist2017}.
In the present work, we have modified the cold field emission source to integrate optics in the immediate vicinity of the field emission cathode to minimize the size of the laser focal spot  on the tip apex, minimize the size of the emission region and therefore optimize the brightness of this new kind of ultrafast TEM electron source.
Furthermore, CFE sources do not need to be heated to operate in continuous (DC) emission and therefore the proposed architecture can operate either in conventional DC or ultrafast mode and switching between the two modes is easily done by changing the extraction voltage.
In the following, we present the design of the instrument, its performances and illustrate the potential of the new ultrafast source on a few TEM applications.

\section{Development of an ultrafast Transmission Electron Microscope based on a Cold Field emission source}

\subsection{Accessing temporal resolution in TEMs}

\begin{center}
\begin{figure}[htp]
\centering
\includegraphics[width=12cm,angle =0.]{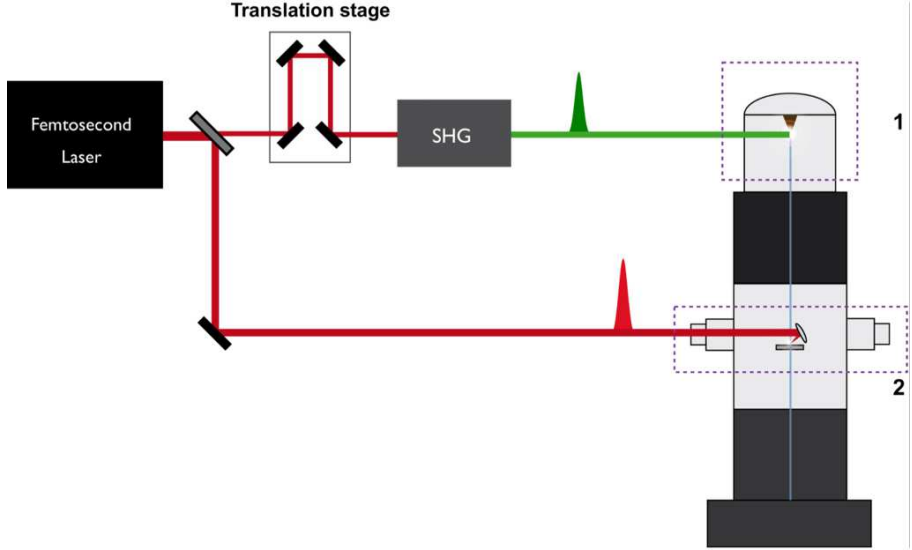}
\caption{(Color Online) Schematic of a time-resolved pump-probe TEM experiment. The emission of electron pulses is triggered by a laser pulse inside the electron gun of the microscope (1). The objective lens of the latter must be adapted to allow for light injection inside the column to excite the sample (2). SHG: Second Harmonic Generation.}
\label{Figure1}
\end{figure}
\end{center}

Most of ultrafast time-resolved TEM experiments are called pump-probe experiments which involve an optical pulse and a delayed electron pulse \cite{Zewailbook}.
As shown in Figure \ref{Figure1}, the optical \textit{pump} pulse first brings the sample, located in the TEM objective lens, out of equilibrium and the electron \textit{probe} pulse, delayed with respect to the excitation, is used to probe the sample during its relaxation.
By systematically changing the delay between pump and probe, it is possible to record the dynamical evolution of the sample as it goes back to equilibrium.
The delay between pump and probe can be controlled by moving the mechanical delay stage placed on one of the optical paths.
In time-resolved TEM experiments, the temporal resolution depends on the laser pulse duration, electron emission characteristics (initial energy spread, number of electrons per pulse), and propagation inside the TEM (acceleration length, voltage...)\cite{Gahlmann2008}.
As shown in Figure \ref{Figure1}, the generation of the electron probe pulse is triggered by a second optical pulse originating from the same laser source as  the pump pulse and is therefore synchronized with the latter.

Time-resolved TEM experiments can be performed in two different modes.
In the \textit{single-shot mode}, only one electron pulse containing a sufficient number of electrons to yield an exploitable signal (image, diffraction pattern,  spectrum) is used to probe the sample with a delay with respect to the excitation of the sample.
This mode is used to investigate irreversible processes such as phase transitions for instance.
The large number of electrons inside each electron bunch can have a detrimental effect on the temporal and spatial resolution\cite{Gahlmann2008}.
Whereas flat photocathodes can yield electron bunches having more than $10^9$ electrons per pulse in the probe, the maximum number of electrons which can be photoemitted by a metallic nanotip with a single optical pulse lies typically within the range 1-1000 depending on the tip material, laser power, pulse duration, wavelength and repetition rate.
Therefore, laser-driven electron sources based on field emitters cannot be used in the single shot mode\cite{Hilbert2009}.
Furthermore, as will be shown later, at 1 MHz repetition rate, the number of electrons emitted per laser pulse by a metallic nanotip translates into a maximum current which lies between 0.16-160 pA.
This estimation, which ignores current losses between the nanotip and the sample, shows that UTEMs based on laser-driven nanoemitters operate in low-dose and that useful data can only be obtained by accumulating electrons from a sufficiently large number of pulses, in the so-called \textit{stroboscopic observation mode}.
From now on, we will only refer to \textit{probe} currents measured inside the TEM column.

The small number of electrons in each pulse yields an excellent spatio-temporal resolution.
However, this mode of investigation can only be used to investigate reproducible phenomena.
The UTEM developed in the present work is based on a laser-driven cold field emission source and will be used for stroboscopic pump-probe investigations.

\subsection{General presentation of the modified ultrafast coherent TEM}

\begin{center}
\begin{figure}[htp]
\centering
\includegraphics[width=12cm,angle =0.]{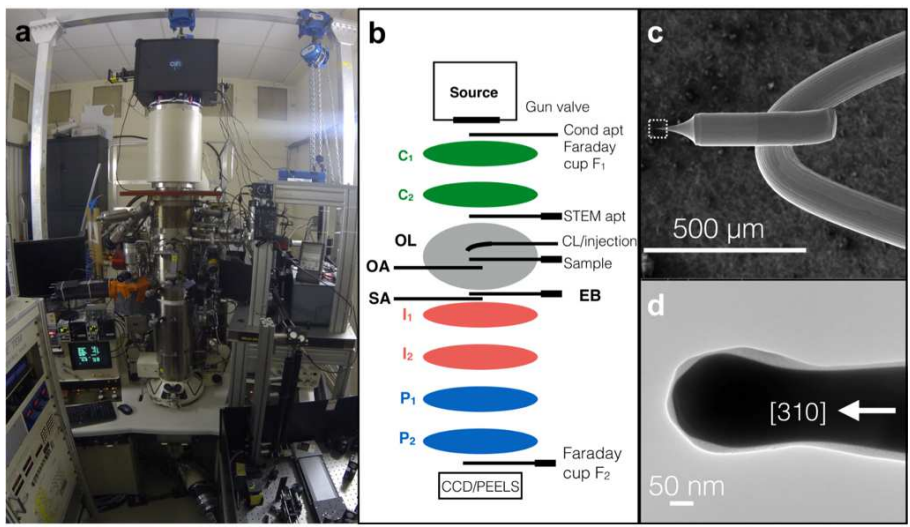}
\caption{(Color Online) a) Picture of the modified HF2000. The black box on top is the optical head designed to steer a femtosecond laser beam onto the apex of the field emission tip inside the electron gun. b) Schematic of the TEM column. $C_1$ and $C_2$: first and second condensor lenses. $OL$: objective lens. $OA$: objective aperture. $EB$: electron biprism. $SA$: selected-area aperture. $I_1$ and $I_2$: intermediate lenses. $P_1$ and $P_2$: projector lenses. c) Scanning Electron Micrograph (SEM) of a commercial HHT [310] oriented tungsten nanotip. d) TEM image showing the (310) crystal facet of the tungsten tip.}
\label{Figure2}
\end{figure}
\end{center}

The present development has been achieved on an \textit{Hitachi High Technologies} (HHT) HF2000 TEM fitted with a 200 kV cold field emission gun.
Figure \ref{Figure2}-a shows a picture of the modified ultrafast TEM.
The black box on top, termed \textit{optical head} in the following, allows to align and control the position of a femtosecond laser beam inside the electron gun.
Details on the optical head are provided later.
The optical head is rigidly fixed onto a metallic housing called \textit{gun-housing}.
The latter is filled with $SF_6$ to electrically insulates the high voltage parts of the electron gun from the earth potential.
Part of the optical set-up (mechanical delay stage) is visible on  the optical table in the bottom right of Figure \ref{Figure2}-a.
A secondary optical breadboard has been added close to the objective lens.
It supports an optical system to align and focus the pump beam inside the objective lens and a spectrometer to analyze cathodoluminescence signals.
Figure \ref{Figure2}-b gives the details of the TEM column.

The TEM column is basically unchanged below the extraction anode (located before the accelerating tube) compared to the original design except in the objective lens (see part 2.4).
The latter has been modified to include a light injection/cathodoluminescence system and an inspection CCD camera to control the insertion of the parabolic mirror inside the objective lens.
The modified HF2000 is fitted with two faraday cups used to determine the probe current, one at the exit of the FE source (called \emph{Faraday cup} $F_1$ in the following), and one above the viewing screen (\emph{Faraday cup} $F_2$).
The two Faraday cups are connected to a picoammeter to monitor the emission in ultrafast mode.
The UTEM is fitted with one electrostatic M\"{o}ellensted biprism above the intermediate 1 lens, a Gatan 4kx4k USC1000 camera and an electron energy loss spectrometer Gatan PEELS666. 
A SEM and a TEM image of a typical HHT commercial [310] oriented tungsten tip such as those used throughout this study are provided respectively in Figure \ref{Figure2}-c and Figure \ref{Figure2}-d.

\subsection{Pump-probe Optical Set-up, TEM Optical Head and Modified Gun Assembly}

Our laser system is based on a compact ultrafast fiber laser, delivering ultrashort (350 fs), high energy (up to
20 $\mu$J), high repetition rate pulses at 1030 nm (Satsuma, Amplitude Syst\`{e}mes).
Electron emission has been obtained using the femtosecond pulses generated at 515 nm by Second Harmonic Generation (SHG) of the laser output in a beta barium borate (BBO) crystal.
The repetition rate of the laser is 1 MHz unless otherwise specified.
The laser system, frequency conversion and pump-probe set-ups are installed on an optical table close to the TEM.
The laser beam used to trigger electron emission from the tip is first sent onto the secondary optical breadboard using one of the two periscopes visible on Figure \ref{Figure2}-a and then to the optical head.
A telescope is inserted on the path of the beam to optimize its focusing on the tip apex.

\begin{center}
\begin{figure}[htp]
\centering
\includegraphics[width=12cm,angle =0.]{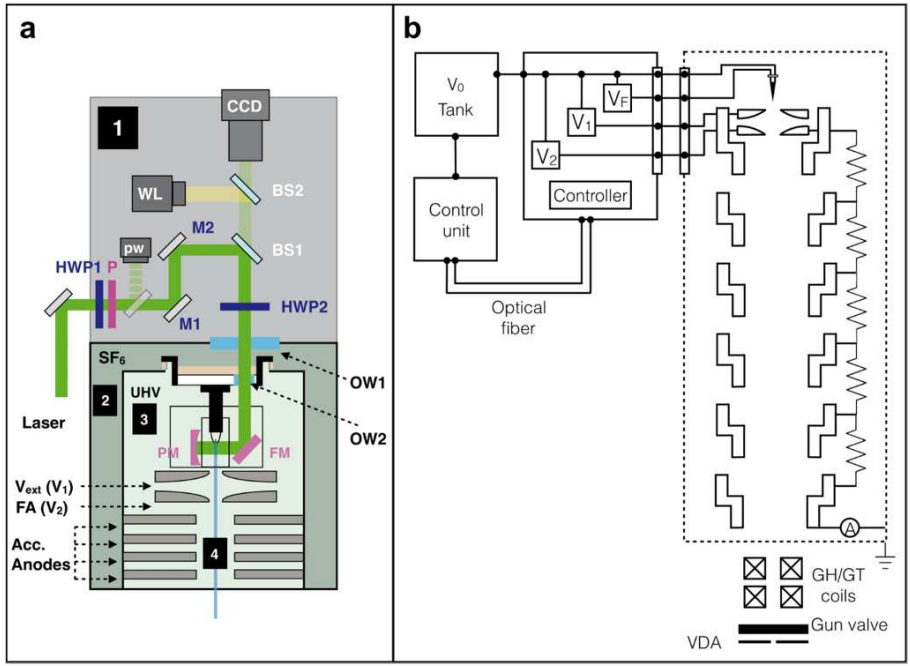}
\caption{(Color Online)  a) Details of the optical head. $HWP$: half-wave plate. $BS$: beam-splitter. $WL$ white-light source. $OW$: optical window. $FM$ flat mirror. $PM$: parabolic mirror. b) High voltage configuration of the HHT HF2000 TEM electron gun.  $V_0$ is the acceleration voltage, $V_1$ is the extraction voltage, $V_2$ the Focusing Anode (FA) voltage and $V_F$ the voltage applied to the tip during the flash-cleaning. GH: gun horizontal. GT: gun tilt. VDA= vacuum differential aperture.
The anodes below $V_2$ are located inside the accelerating tube and the dotted rectangle stands for the gun-housing grounded at the earth potential.}
\label{Figure3}
\end{figure}
\end{center}

Figure \ref{Figure3}-a provides the details of the optical head.
The latter has been designed to allow controlling the laser power, polarization and position on the tip remotely.
The laser beam goes first through an attenuator composed of a half-wave plate and a polarizer and then onto two mirrors $M_1$ and $M_2$ mounted on piezo-driven positioning mirror mounts.
The polarization of the laser is then controlled by a second half-wave plate.
The laser beam enters the  $SF_6$ region inside the gun-housing through a first optical window.
It then enters the ultra-high-vacuum (UHV) of the electron gun through a second optical window.
The latter is included in a DN 40 CF flange which has been modified from the original Hitachi design (see ref. \cite{Caruso2017} for more details).
The laser is then reflected first on a planar mirror and finally on a parabolic one.
The focal distance of the latter is $f'=8$ mm which yields a focal spot diameter ($1/e^2$ diameter) of $\sim$ 3 $\mu m$.
The design of the two mirrors and their holder, their compatibility with the UHV and high voltage requirements of the electron gun have been carefully optimized.
In particular, the mirror holder has been mirror polished as it is located at less than 2 mm from the extraction anode  (surface rugosity of $R_a = 0.1 \mu$m).
As will be detailed later, apart from the region of the objective lens, there is no modification of the TEM column below the extraction anode to keep the performances of the instrument as close as possible to the ones of the original HF2000 operating in conventional DC mode.
Care has also been taken to check that the baking procedure (350$^{\circ}$C) to reach UHV condition (around $10^{-9}$ Pa) close to the tip, can also be used with the modified electron source.

\begin{center}
\begin{figure}[htp]
\centering
\includegraphics[width=12cm,angle =0.]{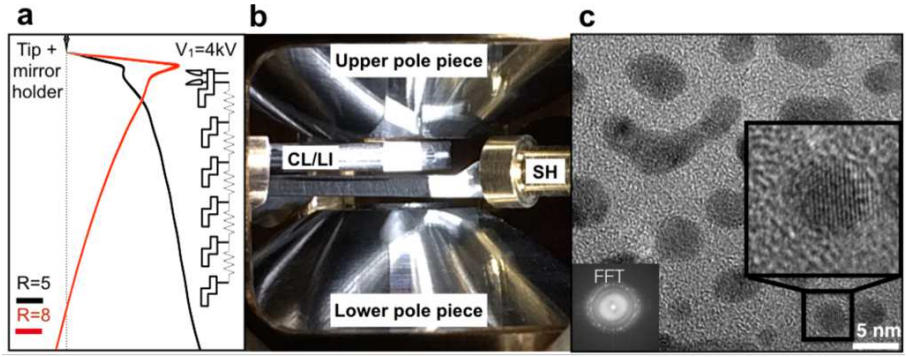}
\caption{(Color Online) a) Electron trajectories inside the ultrafast electron source computed using SIMION for two different values of the voltage ratio $R= V_2/ V_1$. 
b) Image of the objective lens pole piece gap (width = 4.5 mm) showing the cathodoluminescence/light injection (CL/LI) system inserted above the sample holder (SH).
c) High Resolution TEM image obtained with the modified TEM operated in continuous (DC) emission mode on gold nanoparticles confirming that no significant deterioration of the spatial resolution is introduced by the modifications performed to allow ultrafast operation of the TEM.}
\label{Figure4}
\end{figure}
\end{center}

The high voltage configuration of the HF2000 electron source is described in Figure \ref{Figure3}-b. 
The tip and the mirror holder system are set at the acceleration voltage $V_0$ (usually -200kV) and face the extracting anode set at a voltage $V_1$ relatively to the tip. 
The distance between the tip and the anode is adjusted in a range of about 7-10 mm. 
The first anode of the accelerating tube, called \textit{focusing anode} or gun lens, is set at a potential $V_2$ relatively to the tip. 
 The other anodes located in the accelerating tube are used to equally distribute the acceleration voltage between the focusing anode and the earth potential (around 33 kV per stage for 200 kV total acceleration voltage).
The ratio $R= V_2/ V_1$ between the focusing anode and the extracting anode can be adjusted between 2 and 13.
The ratio is used to set the position of the full gun cross-over which may be either real (which is the condition selected for TEM applications to maximise the beam intensity) or virtual (in order to maximise the first illumination lens demagnification for Scanning TEM (STEM) applications \cite{HHTpatentHF2000}).  
As will be detailed later, when operating in ultrafast mode, the electron emission is triggered by the laser pulses and $V_1$ becomes an adjustable additional parameter which can be used to optimize the electron beam.
The electronics used to flash-clean the nanotip has not been modified.
As already described in ref. \cite{Caruso2017} SIMION and EOD  (Electron Optics Design software) simulations of the new ultrafast source show that the mirrors and mirror holder, inserted close to the tip, do not alter the electron optics properties of the source. 
Figure \ref{Figure4}-a shows two examples of electron trajectories computed inside the modified electron gun using SIMION 8.1 for two different ratios yielding either a real or a virtual cross-over\cite{Dahl2000}.
To assess the influence of the insertion of additional components in the vicinity of the field-emission tip, we have computed the aberrations of the ultrafast electron source using EOD.
In real cross-over condition the spherical and chromatic aberration coefficients of the ultrafast CFEG are respectively $C_s \approx$ 60mm and $C_c \approx$ 12mm while in virtual cross-over condition they are respectively $C_s \approx$ 32mm and $C_c \approx$ 10mm which are in the same range as the ones of the original gun.
Aberrations affect the spot size formed by the gun and are one of the major brightness limiting factor of CFE sources.
To achieve a nanometric electron probe on the sample plane, the first illumination lenses, located after the gun, are used to strongly demagnify this first spot size. 
State of art (S)TEM use a spherical aberration corrector, which enables to cancel the spherical aberration contribution of the illumination lenses system and to generate bright sub-Angstr\"{o}m probe.
To confirm that the modification of the electron gun has a negligible impact on the TEM performances, HREM images have been acquired in conventional DC mode on gold nanoparticles yielding an image lattice resolution of 0.2 nm at 200kV of acceleration voltage (see Figure \ref{Figure4}-c).

\subsection{Light injection and collection from inside the objective lens}

The objective lens of the HF2000 has been modified to allow injection, focusing and precise alignment of an ultrafast laser beam onto the sample. 
The injection/CL system is composed of a parabolic mirror placed above the sample holder inside the 4.5 mm narrow pole-piece gap of the HF2000 and mounted on an XYZ translation stage with differential micrometric screws\cite{kociakpatent2013}.
Figure \ref{Figure4} shows the injection/CL system inserted above the sample holder inside the pole piece gap.
An optical window sealed on a tube allows a laser beam propagating in free space to enter inside the TEM column and be focused on the sample.
The parabolic mirror has a focal distance of 1 mm and a hole to let the electrons reach the sample.
It can be inserted or completely removed from above the sample in the objective lens using a mechanical translation stage.
An inspection CCD has been mounted on the standard cold trap port to monitor insertion and removal of the parabolic mirror and avoid mechanical interferences between the light injection system and the sample holder.
As shown in Figure \ref{Figure2}-a, an optical breadboard has been installed in the vicinity of the objective lens.
It includes the optical set-up used to align the pump beam inside the objective lens, a spectrometer and a CCD camera for cathodoluminescence experiments.
Mirrors mounted on piezo-actuated mounts are also used to optimize the alignment of the laser beam on the sample inside the objective lens.
For cathodoluminescence experiments, a bundle of optical fibers is inserted inside the tube in close proximity of the optical window to maximize the light collection efficiency.
A spectrometer coupled to a sensitive CCD camera and photodetectors are available for low-level light detection.
Laser injection on the sample and cathodoluminescence experiments have been performed in the TEM DC emission mode confirming that the system is fully operational. 
In the following, we only discuss the potential of the new ultrafast electron source for TEM using ultrashort coherent electrons pulses.


\section{Optimization and Characterization of the Ultrafast Electron Probe}

\subsection{Optical Alignment Procedure, Laser Power and Polarization dependencies of the Ultrafast Electron Probe}

\begin{center}
\begin{figure}[htp]
\centering
\includegraphics[width=12cm,angle =0.]{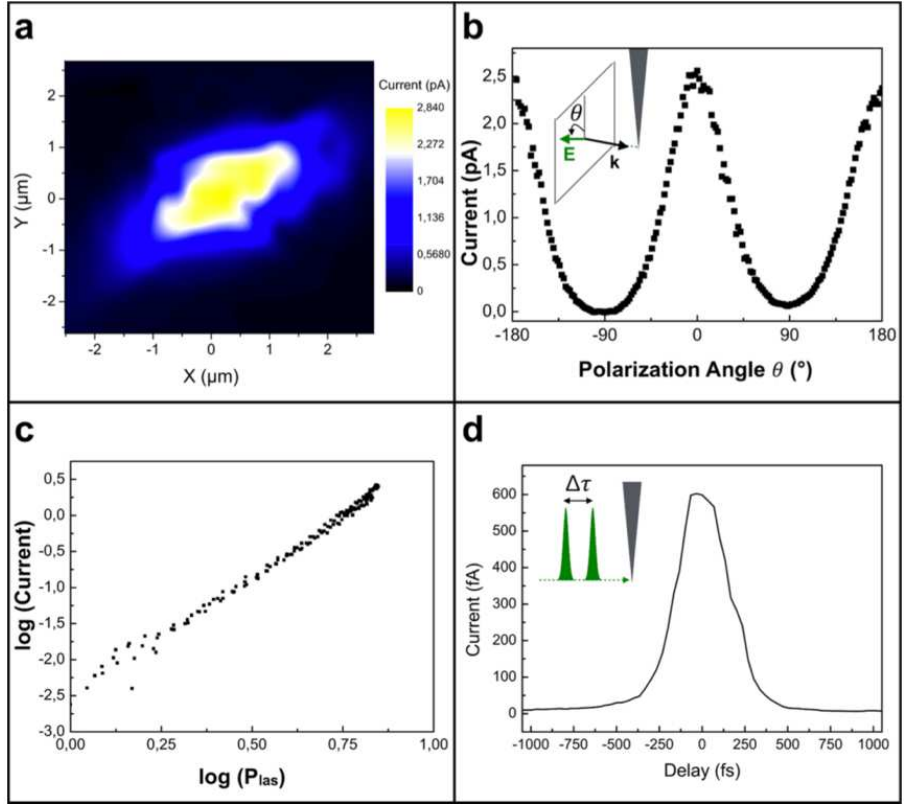}
\caption{(Color Online) a) map of the probe current measured by the Faraday cup ($F_1$) obtained by scanning the position of the laser beam on the tip apex using $M_2$ (see Figure \ref{Figure2}-a). b) Probe current measured by the Faraday cup ($F_1$) for different polarizations of the femtosecond laser pulses focused on the nanotip inside the electron gun. c) Dependence of the ($F_1$) probe current on the incident laser power. d) Two-pulse correlation measurements of the ($F_1$)  probe current.}
\label{Figure5}
\end{figure}
\end{center}

Figure \ref{Figure5}-a is a map of the probe current measured by \textit{Faraday 1} ($F_1$) obtained by scanning the position of the laser beam on the tip using $M_2$. 
It clearly shows that electron emission is restricted to a region having an extension smaller than 1 $\mu$m.
The maximum probe current measured is 2.8 pA in this case with an incident laser power of 4.5 mW and a repetition rate of 1 MHz.
This corresponds to an average of $\sim$ 20 electron per laser pulse inside the probe beam.
Assuming that the electron emission is restricted to the 250 fs of the laser pulse (see Figure \ref{Figure5}-d), this instantaneous probe current would correspond to a DC probe current of 11 $\mu$A.
Figure \ref{Figure5}-b shows the polarization dependence of the $F_1$ probe current: maximum emission is obtained when the polarization of the electric field of the laser is parallel to the tip axis.
Figure \ref{Figure5}-c shows the dependence of the $F_1$ probe current on the incident power.
Our results are consistent with the literature and show that electron emission is triggered from the apex of the nanotip by the multiphoton photoemission.
Figure \ref{Figure5}-d shows the results of two-pulse correlation measurements on the $F_1$ probe current.
These measurements consist in sending a sequence of two laser pulses on the nanotip and measuring the probe current as a function of the delay between the two pulses.
They show electron emission is restricted to a short time window of less than 400 fs.
These results show that cumulative heating effects can be discarded.
This is confirmed by an estimation of the average temperature rise inside the nanotip apex which is of the order of 10 K.  
 
 When performing a cold start of the UTEM, after flash-cleaning the tip and ramping the high voltage, a scan of the laser beam such as the one shown in 
 Figure \ref{Figure5}-a is first realized.
 The position of the laser beam is then finely adjusted and the laser focusing optimized if needed by adjusting the telescope.
 Typically, a laser power of 5 mW yields a probe current in the range 2-3 pA at 1 MHz.
 A cold start of the TEM in ultrafast mode takes about 30 min before the UTEM can be used for TEM applications such as the ones presented in the next part of this article. 

\subsection{Long-Term and Short-Term Stability of the Ultrafast Electron Probe}

\begin{center}
\begin{figure}[htp]
\centering
\includegraphics[width=8cm,angle =0.]{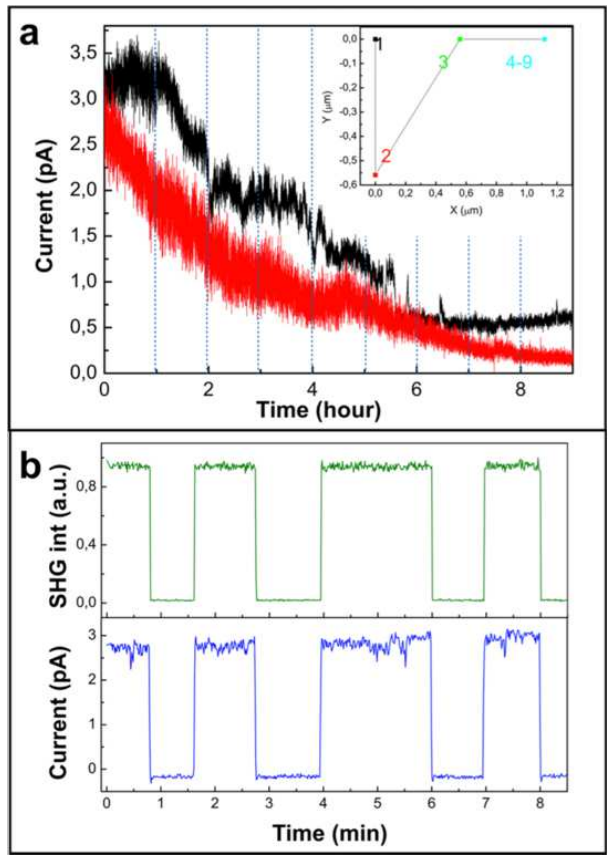}
\caption{(Color Online) a) Long-Term stability of the ultrafast electron probe current measured during 9 hours. The nanotip has been flashed once at the
beginning of the acquisition. Red: no optimization of the laser beam position during the 9 hours. Black: laser beam position has been reoptimized every hour. 
b) Short-term stability of the ultrafast electron probe and simultaneous measurement of the SHG intensity.}
\label{Figure6}
\end{figure}
\end{center}

Figure \ref{Figure6}-a shows data on the long-term stability of the ultrafast probe current measured with the Faraday cup $F_1$ over 9 hours without any flash-cleaning of the tip during the measurement. Figure \ref{Figure6}-a corresponds to the ultrafast CFE source set with an initial probe current of 2.5 pA generated by a laser power P = 7.1 mW at a repetition rate of 1 MHz, an extraction voltage $V_1$=4 kV and a ratio of 5. 
The same measurement obtained on a standard continuous (DC) emission CFE source can be found in ref. \cite{Mamishin2017}.
As for standard CFE source, the ultrafast probe current decreases due to tip surface contamination \cite{Crewe1968,Swanson2009}. To retrieve the original probe current value, the tip needs to be flash-cleaned. In ultrafast emission, in addition to tip contamination, the probe current decrease visible in Figure \ref{Figure6}-a could be due to a slow misalignment of the laser spot onto the tip apex over the 9 hours of measurement. To discriminate the two contributions, the same acquisition has been repeated over 9 hours, with a re-optimization of the laser position every hour. As clearly shown in Figure \ref{Figure6}-a the probe current is still decreasing demonstrating that the major contribution comes from the tip surface contamination as in the case of standard CFEG. 
The inset of Figure \ref{Figure6}-a shows that during the last 5 hours the laser position yielding the highest probe current did not change confirming the good stability of the optical set-up.
The observed decrease in probe current could be improved using modern ultra high vacuum systems and new flash-cleaning technologies as those recently presented in various manufacturers microscopes\cite{Kasuya2010,Cho2013}.

The amplitude of the high frequency current fluctuations, usually called tip noise, is larger than in standard CFEG with a relative standard deviation of 8$\%$ determined in Figure \ref{Figure6}-a compared to 1$\%$ usually observed in DC emission. In DC cold field emission, tip noise generally arises from three main contributions \cite{Swanson2009}: adsorption/desorption processes, thermally induced transitions between two bonding states of adsorbed molecule and surface diffusion of adsorbed molecules. The weight of each contribution depends on the vacuum level around the tip. Usually, in UHV the surface diffusion effect dominates. In ultrafast emission, the tip noise could also be induced by fluctuations of the SHG intensity reaching the optical head. In order to discriminate all contributions, we have performed a simultaneous measurement of the probe current noise and the SHG intensity. Figure \ref{Figure6}-b reports the two curves showing that the fluctuations of the SHG intensity (around 2$\%$) cannot alone explain the detected noise. Furthermore, we can clearly see that the current noise is still important when the laser is blanked \textit{i.e} when the probe current stops. So the noise behaviour is very likely the result of the combined influence of standard contamination, surface diffusion on the tip surface, fluctuations of the SHG intensity and the detection noise. 

\subsection{Brightness and Angular Current Density of the Ultrafast Electron Probe}

To evaluate the potential of the new ultrafast transmission electron microscope (TEM), we have first characterized the electron probe.  

As described previously, in ultrafast mode, the laser beam triggers electron emission by multiphoton photoemission, the applied extraction voltage $V_1$ being kept at a level too low to induce DC emission. This means that, in ultrafast emission condition, the extraction voltage $V_1$ can be used as a free parameter to adjust the strength of the electrostatic gun lens if its value remains below the onset of DC emission.

\begin{center}
\begin{figure}[htp]
\centering
\includegraphics[width=10cm,angle =0.]{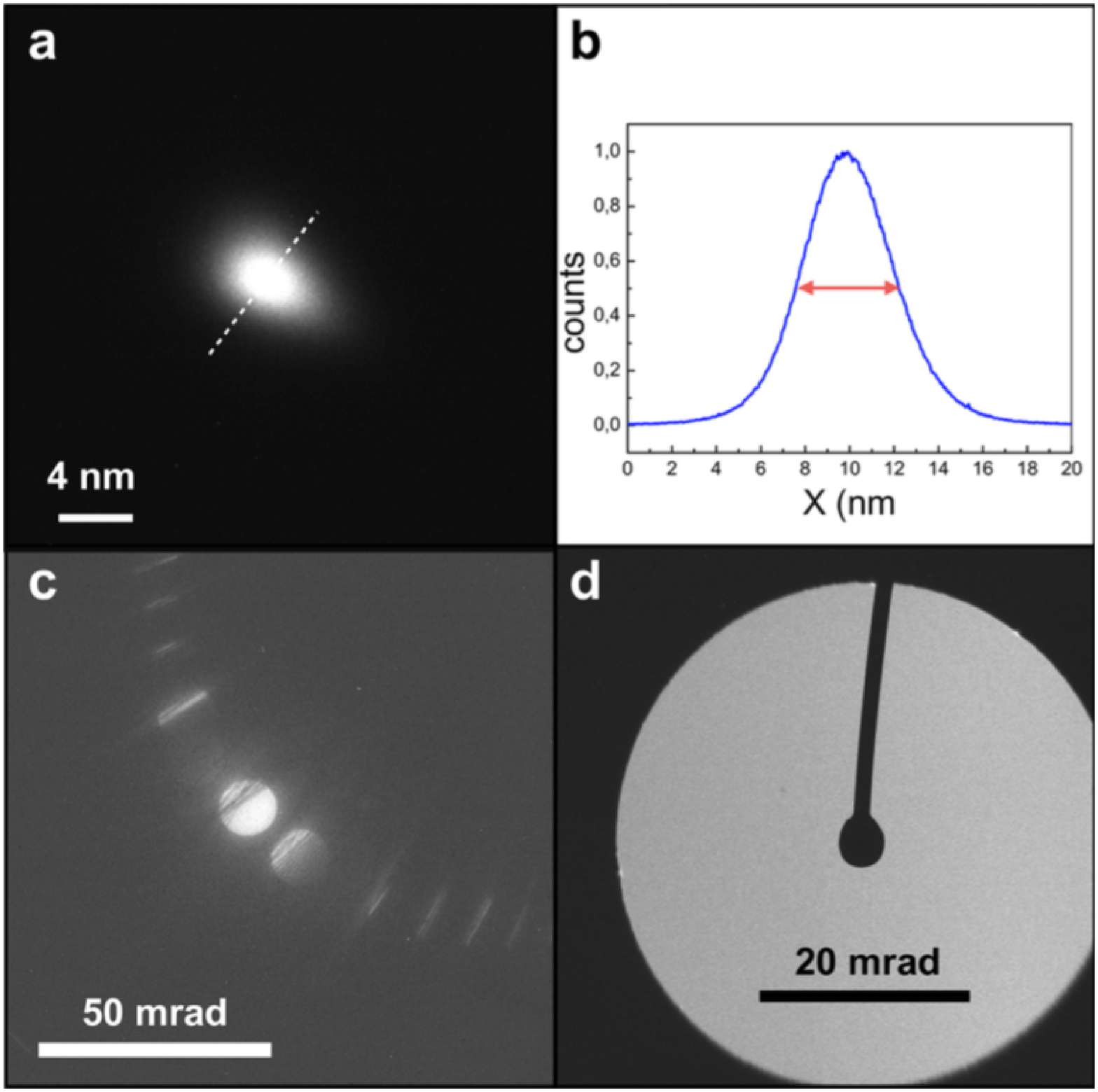}
\caption{(Color Online) Experimental measurement of the brightness in virtual cross-over mode. a) 4 nm diameter (fwhm) spot obtained using a 30 $\mu$m STEM aperture. b) Profile of the electron spot along the dotted line defined in \ref{Figure7}-a. c) CBED pattern of a Si sample oriented in (220) two beam condition used to calibrate the convergence angle. 
d) Transmitted CBED disk obtained in diffraction and analysis mode of the microscope with a 100 $\mu$m STEM aperture used to determine the angular current density of the ultrafast source. The shadow of the Faraday cup $F_2$ is visible in the center.}
\label{Figure7}
\end{figure}
\end{center}

The brightness in ultrafast emission mode has been determined both in real and virtual cross over conditions of the gun. 
Figure \ref{Figure4}-a reports the associated electrons trajectories simulated by SIMION \cite{Dahl2000,Kubo2017}.
Measurements were performed in "analysis" mode of the microscope, used to maximise the demagnification power of the illumination system and routinely chosen for CBED (Convergent Beam Electron Diffraction) and STEM applications (\textit{i.e} equivalent to the so-called "nanoprobe" or "CBD" modes for FEI and JEOL microscopes respectively). 
A 30 $\mu$m STEM aperture has been used to minimize the influence of condensor spherical aberration on the measurement.  
As reported in Figure \ref{Figure7}, in virtual cross-over mode ($V_1$=4 kV and R=5), for a spot radius of 2 nm (taken from the full width half maximum of the spot intensity), a half convergence angle of 6 mrad (calibrated in CBED using the (220) reflexion of a Si sample), a current of 80 fA has been measured with Faraday cup $F_2$. Neglecting the contribution of the illumination system aberrations to the spot size, we obtain a brightness value of 5.8x$10^7$ $\mathrm{A.m}^{-2}\mathrm{.Sr}^{-1}$. Under the same emission conditions (laser power and gun voltages) a probe current of 2.5 pA is measured using the Faraday cup $F_1$ after the gun and before the STEM aperture. The loss of current mainly comes from the global contribution of spherical aberrations (gun + illuminations lenses). 
Even if we take into account the fact that this measurement has been performed at 1 MHz instead of 250 kHz in previous studies, the measured value is the highest reported on UTEMs based on laser-driven nanoemitters\cite{Feist2017}.
It is worth noting that it has been obtained using the HF2000, a 30 year-old microscope and will be strongly improved using state-of-the-art electron optical instruments. 
In the real cross over conditions ($V_1$ = 4 kV and R = 8), for a spot radius of 4 nm (taken from the full width half maximum of the spot intensity), a half convergence angle of 6 mrad, a current of 122 fA has been measured using Faraday cup $F_2$. 
Neglecting the condensor aberrations, we obtain a brightness value of 2.2x$10^7$ $\mathrm{A.m}^{-2}\mathrm{.Sr}^{-1}$.
At a 1 MHz laser repetition rate, electron emission occurs during $\sim$ 250 fs, \textit{i.e} the laser pulse duration, every 1 $\mu$s.
Therefore, assuming a constant value of the instantaneous current, the measured value would correspond in DC to a brightness in the $10^{13}$ $\mathrm{A.m}^{-2}\mathrm{.Sr}^{-1}$, \textit{i.e} similar to state-of-the-art DC cold-field emission sources.
This confirms that the ultrafast electron source maintains its unique brightness even in ultrafast mode allowing to generate a high spatial coherence beam.
Stated differently, our results show that by inserting the focusing optics inside the UHV and high voltage environment of the electron source as close as possible to the cathode and thereby minimizing the size of the emission region, it is possible to keep the brightness of the electron source unaltered.
The major counterpart of the ultrafast mode is the decrease in current.

Figure \ref{Figure7}-d reports also the experimental results of the angular current density measurement performed in the analysis  and diffraction mode with a 100 $\mu$m STEM aperture. The Faraday cup $F_2$ is used to collect the current in a given solid angle taken in the center of the transmitted disk. Using such a configuration we have determined an angular emission density of 2 nA.Sr$^{-1}$.
Our measurements of the brightness and angular current density yield a virtual source size in the 5-10 nm range, which is the smallest achieved to date in UTEMs  and equivalent to standard CFEG virtual source size.

\section{Performances of the Ultrafast Transmission Electron Microscope}

In order to illustrate the capabilities of the new FE-UTEM, conventional TEM imaging as well as electron diffraction in parallel (SAED, Selected Area Electron Diffraction) and convergent beam configuration (CBED), Electron Energy Loss Spectrometry (EELS) and electron holography have been performed using ultrashort electron pulses. All results presented below have been obtained at an acceleration voltage of $V_0$=-150 kV. 

\subsection{Imaging, diffraction and spectroscopy with ultrashort electron pulses}

\begin{center}
\begin{figure}[htp]
\centering
\includegraphics[width=12.5cm,angle =0.]{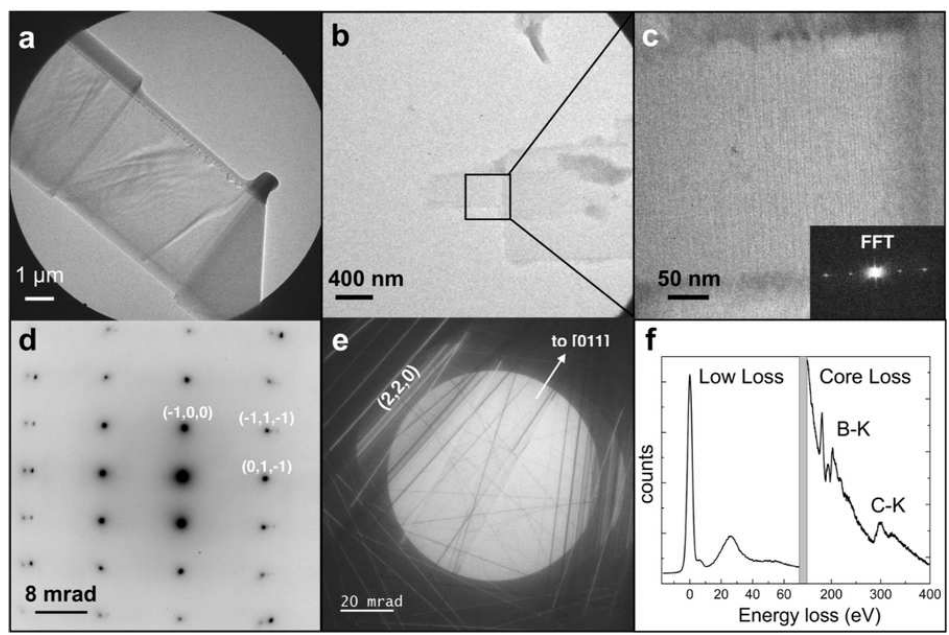}
\caption{(Color Online) a) ultrafast conventional TEM image of a Si lamella. 
b) and c)  ultrafast HREM image of a biological Catalase crystal.
d) ultrafast SAED pattern taken along the [110] direction of a TiAl $\gamma$-phase crystal. e) Ultrafast CBED pattern obtained near [110] direction of a Si crystal. f) Ultrafast EELS spectrum of a boron nitride crystal.}
\label{Figure8}
\end{figure}
\end{center}

Figure \ref{Figure8}-a and \ref{Figure8}-b-c report respectively an ultrafast conventional TEM image of a Si lamella and an ultrafast high-resolution image of a biological Catalase crystal. Figure \ref{Figure8}-d and \ref{Figure8}-e report respectively an ultrafast SAED pattern taken along the [110] direction of a TiAl $\gamma$-phase crystal and the ultrafast CBED pattern obtained near [110] direction of a Si crystal. Resolutions of the ultrafast images and diffraction patterns are comparable to the one usually obtained with continuous HF2000 FE-TEM. Lenses conditions remained unchanged for all observations mode of the TEM as the modifications operated inside the gun are located before the extraction anode.
Furthermore, no perturbations have been observed due to the presence of the parabolic mirror inside the objective lens pole piece gap. This is confirmed by the high resolution images of gold nanoparticles obtained in DC emission mode with all the injection system inserted as previously reported in Figure \ref{Figure4}-c. Indeed, aberrations of the objective lens remain unchanged and are respectively $C_s$=1.2 mm and $C_c$=1.4mm. 
In ultrafast mode, under parallel illumination mode we achieved an ultimate lattice image resolution of 0.9 nm at 150kV measured thanks to a Crocidolite crystal. We were unable to resolve gold atomic structure as for the DC emission case, due to the mechanical stability of the TEM under very high exposure time (150 s in this case) coming from the limited probe current.  In convergent illumination mode, we achieved an ultimate spot size of 1.5 nm with 20 fA of probe current obtained for a $V_1$=2 kV and R = 8. 

As noted before, the major difference with the observation performed in DC emission lies in the amount of emission and probe current. In standard DC mode the emission current is usually set around 10 $\mu$A, which can generate a probe current (in analysis mode using 30 $\mu$m STEM aperture) of approximately 100 pA. In the same electron optical condition ($V_1$=4 kV and R=5), using the ultrafast mode, the emission current is set around 2.5 pA at P = 6 mW and 1 MHz of laser power and repetition rate (approximately 15 electrons per pulse), which can generate a probe current of 80 fA. This value is mainly limited by the tip withstand and the laser repetition rate. 
Therefore, the exposure time has to be increased and the image resolution remains mainly limited by the microscope stability during the exposure time and the quantum efficiency of the detector. 

Finally, ultrafast EELS has been performed on boron nitride sample. The spectrum, reported in Figure \ref{Figure8}-f, was acquired in virtual cross over condition at an acceleration voltage of 150 kV, a repetition rate of 2 MHz and a laser power of 8.4 mW. The microscope was set in diffraction and analysis mode with a 0.15 m camera length. Usual behavior could be observed in the low-loss (bulk plasmon) and core-loss (boron and carbon $K$ edge) regions, but the spectrum resolution remains strongly limited by the beam energy width. Indeed, depending on the number of electrons per pulse (from 1 to 20), we could measure an energy resolution within the interval of 1 to 2.5 eV. We were unable to reach an energy resolution better than 1 eV whereas an energy resolution of 0.45 eV was measured in DC emission mode under the same electron optical conditions.
Further experiments and simulations are under progress to address this issue and will be the subject of a dedicated study.

\subsection{Electron Holography using Ultrashort Electron Pulses}

\begin{center}
\begin{figure}[htp]
\centering
\includegraphics[width=12cm,angle =0.]{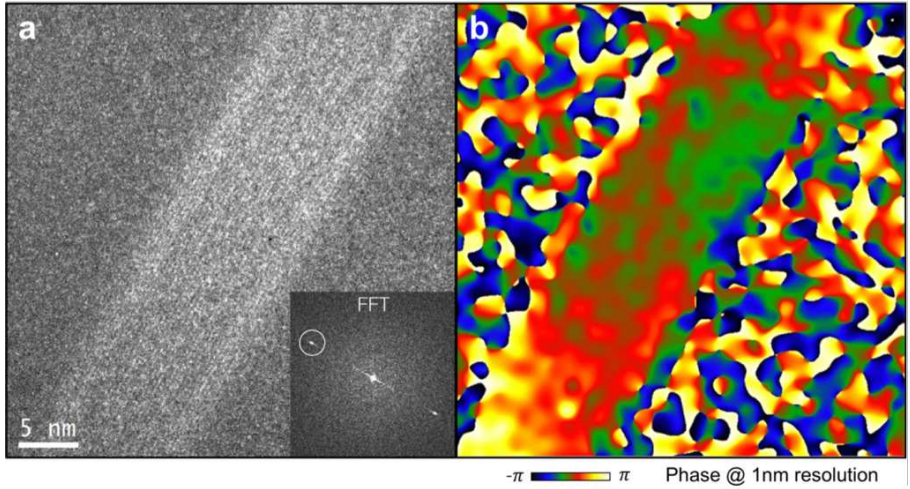}
\caption{(Color Online) a) Ultrafast off-axis electron hologram acquired in vacuum. b) Corresponding phase map obtained at 1nm resolution.}
\label{Figure9}
\end{figure} 
\end{center}

In order to further explore the potential of the UTEM developed in Toulouse for ultrafast electron interferometry, we have acquired ultrafast electron holograms in off-axis configuration. Figure \ref{Figure9}-a reports such a hologram obtained in the vacuum with the microscope sets in "zoom" mode (\textit{i.e} equivalent to so-called microprobe or TEM mode in FEI and Jeol microscope respectively), a 100 $\mu$m condensor aperture, an elliptic illumination perpendicular to the biprism wire \cite{vokl1999}, 50 V of biprism voltage and 150 s of exposure time. The probe current was set at 100 fA under virtual cross over condition ($V_1$= 4 kV and R=5) and 9.5 mW of laser power at 2 MHz of repetition rate. The hologram exhibits 20 $\%$ of fringes contrast with a low signal over noise ratio of 30 counts of mean intensity over 14 counts of noise, originating from the low amount of probe current. Figure \ref{Figure9}-b shows the phase obtained after a standard Fourier filtering of the sideband in the Fast Fourier Transform (FFT) with a 1 nm resolution mask. This result shows that, despite the low signal over noise, the phase information can be retrieved throughout the whole ultrafast hologram width. 
This first ultrafast electron hologram obtained thanks to the high brightness of the ultrafast cold field emission source, paves the way to ultrafast electron interferometry experiments and time-resolved imaging of fields (electric, magnetic, strain) at the nanoscale \cite{Snoeck2008,Hytch2008,Gatel2013}. 

\section{Conclusion}

In conclusion, we have reported about the development of the first ultrafast coherent Transmission Electron Microscope based on a cold-field emission source.
Our results show that the excellent brightness and spatial coherence of cold-field emission sources are maintained in ultrafast mode paving the way towards novel time-resolved ultrafast electron interferometry experiments. 
By inserting the laser focusing optics  close to the cathode tip located inside the UHV and high voltage environment of a cold-field emission source, it has been possible to selectively trigger electron emission only from the apex of the emitter, thereby yielding a virtual source size and brightness beyond previously reported values.
This ultrafast electron source will undoubtedly open new routes towards the time-resolved investigation of electric, magnetic and strain fields in nanoscale objects.
It will also be a unique tool for the time-resolved detection of cathodoluminescence signals from very compact nano-objects assemblies (multiple quantum-wells for instance)\cite{Zagonel2010,Kociak2017}.	
Combining the new coherent ultrafast source and the laser injection system in the objective lens, this instrument will enable studies of novel interaction schemes between electrons and photons such as non-linear electron-photon interactions.

\section{Acknowledgements}

The authors thank the \textit{Institut de Physique du CNRS} and \textit{Agence Nationale de la Recherche} for financial support (ANR grant ANR-14-CE26-0013).
This work was supported by \textit{Programme Investissements d'Avenir} under the program ANR-11-IDEX-0002-02, reference ANR-10-LABX-0037-NEXT (\textit{MUSE} grant).
This work was supported by the computing facility center CALMIP of the University Paul Sabatier of Toulouse.
The authors acknowledge financial support from the European Union under the Seventh Framework Program under a contract for an Integrated Infrastructure Initiative
(Reference 312483-ESTEEM2). The authors are grateful to M. Pelloux for his contribution to the light injector design and fabrication, E. Snoeck and M. J. H\"{y}tch for their support.

\section{References}


\end{document}